\documentclass[prd,aps,preprint,showpacs,nofootinbib]{revtex4-1}

\usepackage{epsfig}
\usepackage{amsmath}

\begin{document}

\title{Non-Universality of Transverse Momentum Dependent Parton
Distributions at Small-$x$}
\author{Bo-Wen Xiao}
\affiliation{Nuclear Science Division, Lawrence Berkeley National Laboratory, Berkeley,
CA 94720}
\author{Feng Yuan}
\affiliation{Nuclear Science Division, Lawrence Berkeley National Laboratory, Berkeley,
CA 94720 and \\
RIKEN BNL Research Center, Building 510A, Brookhaven National Laboratory,
Upton, NY 11973}

\begin{abstract}
We study the universality issue of the transverse momentum dependent
parton distributions at small-$x$, by comparing the initial/final state
interaction effects in dijet-correlation in $pA$ collisions with those in
deep inelastic lepton nucleus scattering. We demonstrate the
non-universality by performing an explicit calculation in a particular model
where the multiple gauge boson exchange contributions are summed up to all
orders. We comment on the implications of our results on the
theoretical interpretation of di-hadron correlation in $dA$ collisions in
terms of the saturation phenomena in deep inelastic lepton nucleus
scattering.
\end{abstract}

\maketitle




\section{Introduction}

The Feynman parton distribution functions describe the internal structure of
hadrons in terms of the momentum distributions of partons in the infinite
momentum frame. These distributions depend only on the longitudinal momentum
fractions of the target hadrons carried by the partons. The measurements of
high energy hadronic processes depending on the Feynman parton distributions
have been made possible by proving the associated factorization theorem,
which guarantees that the parton distributions studied in different
processes are universal~\cite{ColSopSte89}.

In recent years, hadronic physicists become more interested in
semi-inclusive high energy processes, where one can study the intrinsic
transverse momentum of partons inside hadrons. The additional transverse
momentum dependence helps to picture the parton distribution in a
three-dimension fashion and builds the hadron tomograph through the partonic
structure~\cite{Ji:2003ak}. A number of novel hadronic physics phenomena are
also strongly associated with the transverse momentum dependent parton
distributions, such as the single transverse spin asymmetries~\cite%
{BroHwaSch02,Col02,BelJiYua02,{Boer:2003cm}} and small-$x$ saturations
phenomena~\cite{Brodsky:2002ue,{Iancu:2003xm}}. In the last few years,
great progress has been made in understanding the fundamental questions
associated with these transverse momentum dependent parton distributions,
such as the gauge invariance and the QCD factorization~\cite%
{Col02,BelJiYua02,JiMaYu04,{ColMet04}}. In particular, the non-universality
of these distribution functions due to the final/initial state interaction
effects has attracted intensive investigations. It has been found that the
difference between final state interaction in deep inelastic scattering and
the initial state interaction in Drell-Yan lepton pair production in $pp$
collisions leads to an opposite sign for the single spin asymmetries in
these two processes~\cite{BroHwaSch02,{Col02}}. More complicated relation
was further found for the single spin asymmetry in dijet-correlation in $pp$
collisions as compared to those in DIS and Drell-Yan 
processes~\cite{BoeVog03,mulders,qvy-short,Collins:2007nk,
Vogelsang:2007jk,{Rogers:2010dm}}. 
This eventually leads to a conclusion that a standard transverse momentum
dependent factorization breaks down for this process~\cite{Collins:2007nk}.

In this paper, we extend the universality discussions of the transverse
momentum dependent parton distributions to the small-$x$ domain, where the 
$k_{t}$-dependent distributions have been a common practice to describe the
relevant physics phenomena~\cite{Iancu:2003xm}. We expect the
non-universality for these objects as well. However, because of different
approximation has been made in the small-$x$ region, the general arguments
of Refs.~\cite{Collins:2007nk,Vogelsang:2007jk} on the non-universality may
not apply. As far as we know, there has been no discussion on this issue in
the literature\footnote{%
An important factorization breaking effect has been discussed 
in~\cite{Fujii:2005vj}, which is however different from the non-universality issue
we are investigating in this paper. See more detailed discussions below.}.
The objective of this paper is to study this in detail. We will carry out an
explicit calculation in a model where both small-$x$ and low transverse
momentum approximation are valid. Furthermore, we will resum the
initial/final state interactions to all orders in perturbation to study the
associated universality property.

In particular, we investigate the universality of the small-$x$ transverse
momentum dependent parton distributions probed in hadronic dijet-correlation
in nucleon-nucleus collisions, as compared to that in the deep elastic
lepton-nucleus (nucleon) scattering. There have been experimental results on
di-hadron correlation in $dA$ collisions at RHIC reported by the STAR
collaboration, and interesting phenomena were found~\cite{star-data}.
However, the theoretical interpretation is not yet clear at this 
moment~\cite{Qiu:2004da,Marquet:2007vb,Tuchin:2009nf,{Dumitru:2010mv}}. 
As schematically shown in Fig.~1(a), two partons from the nucleon projectile and nucleus
target collide with each other, and produce two jets in the final state,
\begin{equation}
p+A\to \mathrm{Jet1}+\mathrm{Jet2}+X \ ,
\end{equation}
where the transverse momenta of these two jets are similar in size
but opposite to each other in direction. In ideal case, these
two jets are produced back-to-back. However, the gluon radiation
and intrinsic transverse momenta of the initial partons induce the
imbalance between them. We are particularly interested in the
kinematic region that the imbalance $\vec{q}_\perp=\vec{P}_{1\perp}+\vec{P}
_{2\perp}$ is much smaller than the transverse momentum of the
individual jet, namely, $|\vec{q}_\perp|\ll |\vec{P}_{1\perp}|\sim
|\vec{P}_{2\perp}|$. Only in this region, the intrinsic transverse
momentum can have significant effects. Since there are two
incoming partons, both intrinsic transverse momenta can affect the
imbalance between the two jets. For large nucleus and small-$x$,
the dominant contribution should come from the intrinsic
transverse momentum of the parton from the nucleus, for which we
labeled as $q_\perp$ in Fig.~1(a). In the following, we will focus
on this contributions. Of course, we emphasize that both
contributions shall be taken into account to describe the
dijet-correlation in $pA$ scattering.

\begin{figure}[tbp]
\begin{center}
\includegraphics[width=12cm]{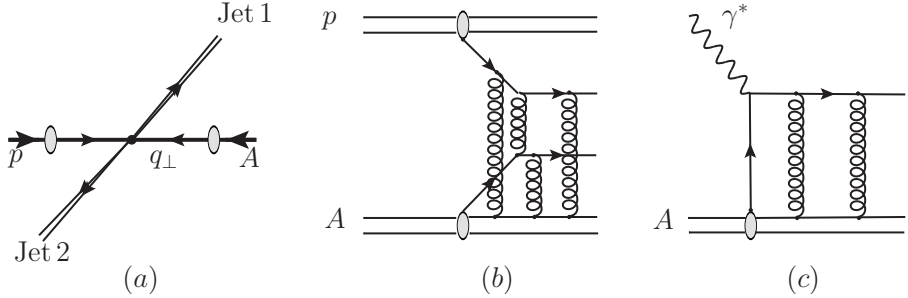}
\end{center}
\caption[*]{(a) Schematic diagram showing that two partons from
the nucleon projectile and the nucleus target collide and produce
two jets in the final state, where the intrinsic transverse
momentum $q_\perp$ from nucleus dominates the imbalance between
the two jets; (b) illustration of initial/final state interactions
which may affect the transverse momentum dependent quark
distribution from the nucleus in this process; (c) as a
comparison, only the final state interaction effect is present in
the deep inelastic lepton-nucleus (nucleon) scattering.}
\label{fig0}
\end{figure}

To understand the universality property of the transverse momentum
dependent parton distribution, we study the
multi-gluon exchange between the hard scattering part with the nucleus 
target~\cite{BelJiYua02,{Brodsky:2002ue},Collins:2007nk,Vogelsang:2007jk}. 
We illustrate the generic diagrams of these interactions in
Fig.~1(b), for the particular partonic channel $qq^{\prime}\to
qq^{\prime}$. All other channels shall follow accordingly. Since
the incoming and outgoing partons are all colored objects, there
exist initial state interaction with the initial parton from the
nucleon projectile, and final state interactions with the outgoing
two partons. For comparison, we also plot in Fig.~1(c) the similar
diagram for the deep inelastic lepton scattering on a nucleus
target, where there is only final state interaction on the struck
quark. Clearly, if the initial/final state interactions affect the
transverse momentum dependence, we will conclude that they are not
universal between these processes. We emphasize, however, that our
discussions will not affect the universality for the transverse
momentum integrated parton distributions. In particular, for the
inclusive observables the initial/final state interaction effects
can be summarized into one gauge link associated with the
integrated parton
distributions, which are universal among different 
processes~\cite{ColSopSte89}.

The rest of the paper is organized as follows. In Sec. II, we
construct a model to investigate the universality property for the
small-$x$ parton distributions, where all gauge boson exchange
contributions can be summed up together, including all initial and
final state interactions. In Sec. III, we summarize our results,
and discuss the phenomenological implications, in particular, on
the theoretical interpretation of the di-hadron correlation in
$dA$ collisions at RHIC recently observed by the STAR
collaboration.

\section{Initial and Final State Interaction Effects}

We take the partonic channel $qq^{\prime}\to qq^{\prime}$ as an example to
show the initial/final state interaction effects and calculate the quark
distribution in dijet correlation $pA\to \mathrm{Jet1}+\mathrm{Jet2}+X$, and
compare with that in the deep inelastic scattering process. At small-$x$,
quark distribution is dominated by gluon splitting, and can be calculated
from the relevant Feynman 
diagrams~\cite{Marquet:2009ca,McLerran:1998nk,{Mueller:1999wm}}. 
For the purpose of our calculation, we employ an Abelian model of 
Refs.~\cite{Brodsky:2002ue,Collins:2007nk,{Vogelsang:2007jk}}. It is a scalar
QED model with Abelian massive gluons with a mass $\lambda$. We
construct the model in such a way that the large nucleus is
represented by a heavy scalar target with mass $M_A$. The scalar
quarks are generated by the Abelian gluon splitting which is the
dominant contribution at small-$x$. The associated quark
distribution in deep inelastic scattering process in this model 
has been calculated in~\cite{BelJiYua02,{Brodsky:2002ue}}.

Since we are interested in studying the final state interaction
effects on the parton distribution of the nucleus, for
convenience, we choose the projectile as a single scalar quark
with charge $g_2$, which differs from the charge of the scalar
quark from the target nucleus $g_1$. In addition, we assume that
the Abelian gluon attaches to the target nucleus with an effective
coupling $g$. All the partons in this calculation is set to be
scalars with a mass $m$. The coupling $g_2$ being different from
$g_1$ is to show the dependence of the parton distribution on the
initial/final state interactions associated with the incoming
parton. If the dependence on $g_2$
remains for the nucleus parton distributions, they are not 
universal~\cite{Collins:2007nk,{Vogelsang:2007jk}}.

\begin{figure}[tbp]
\begin{center}
\includegraphics[width=4cm]{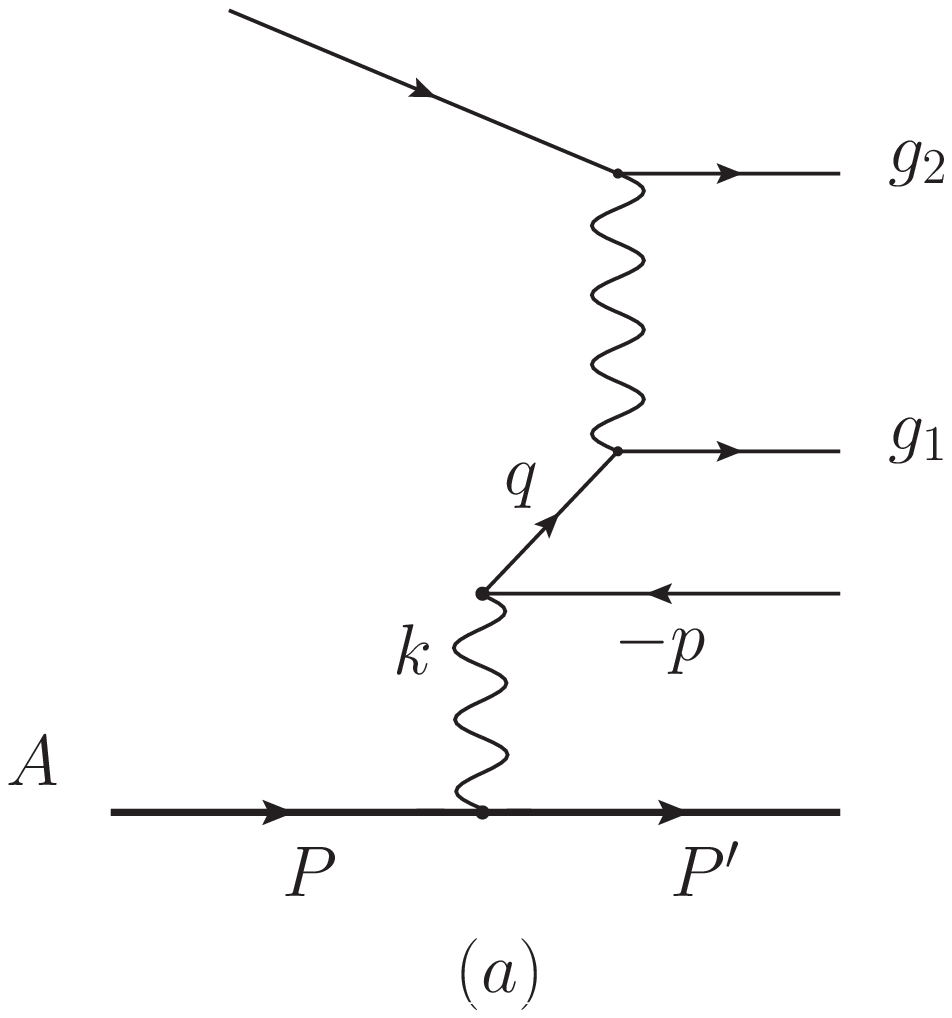}\hfill  \includegraphics[width=4cm]{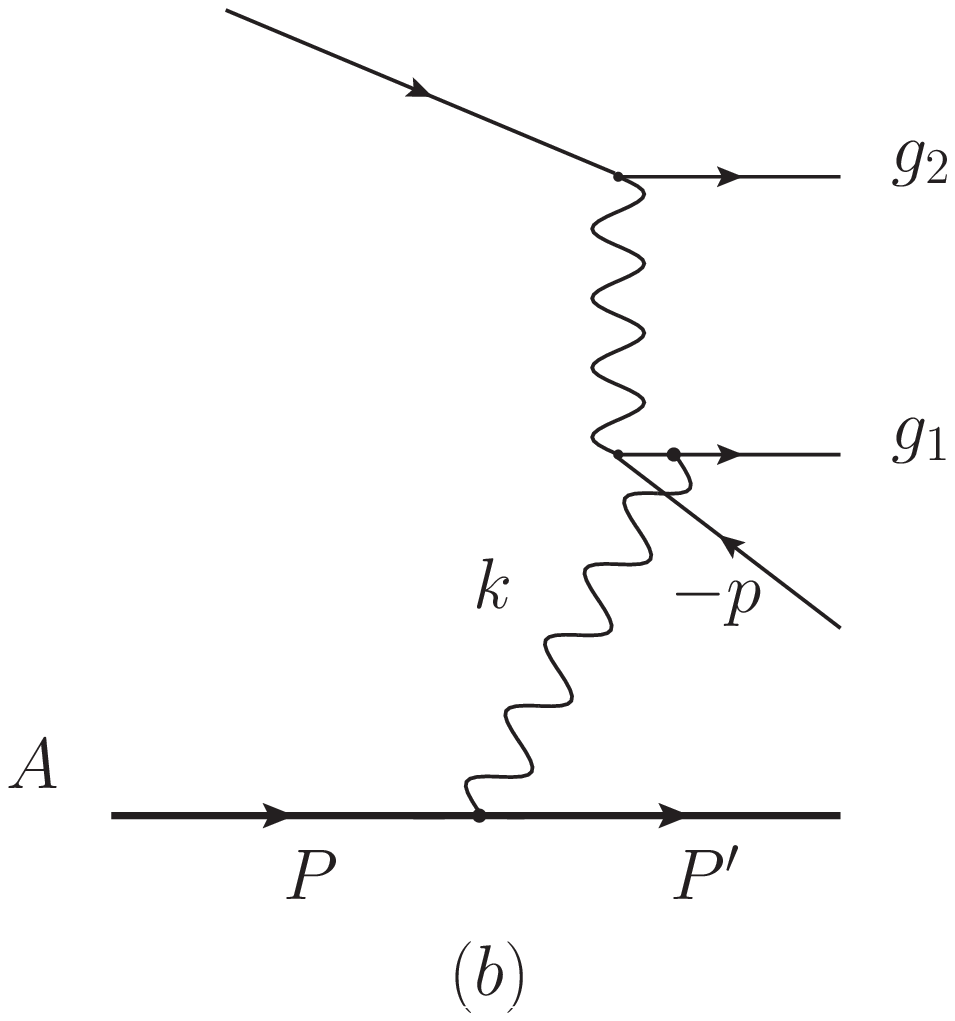}%
\hfill 
\includegraphics[width=4cm]{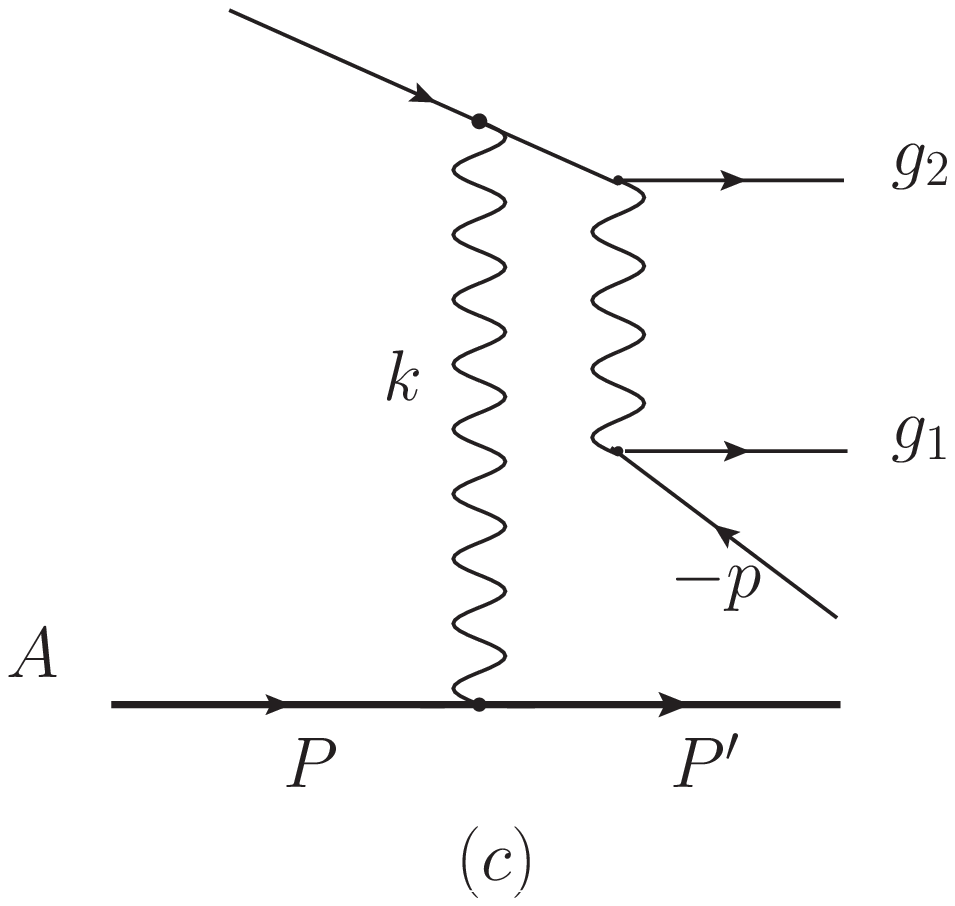}\hfill \includegraphics[width=4cm]{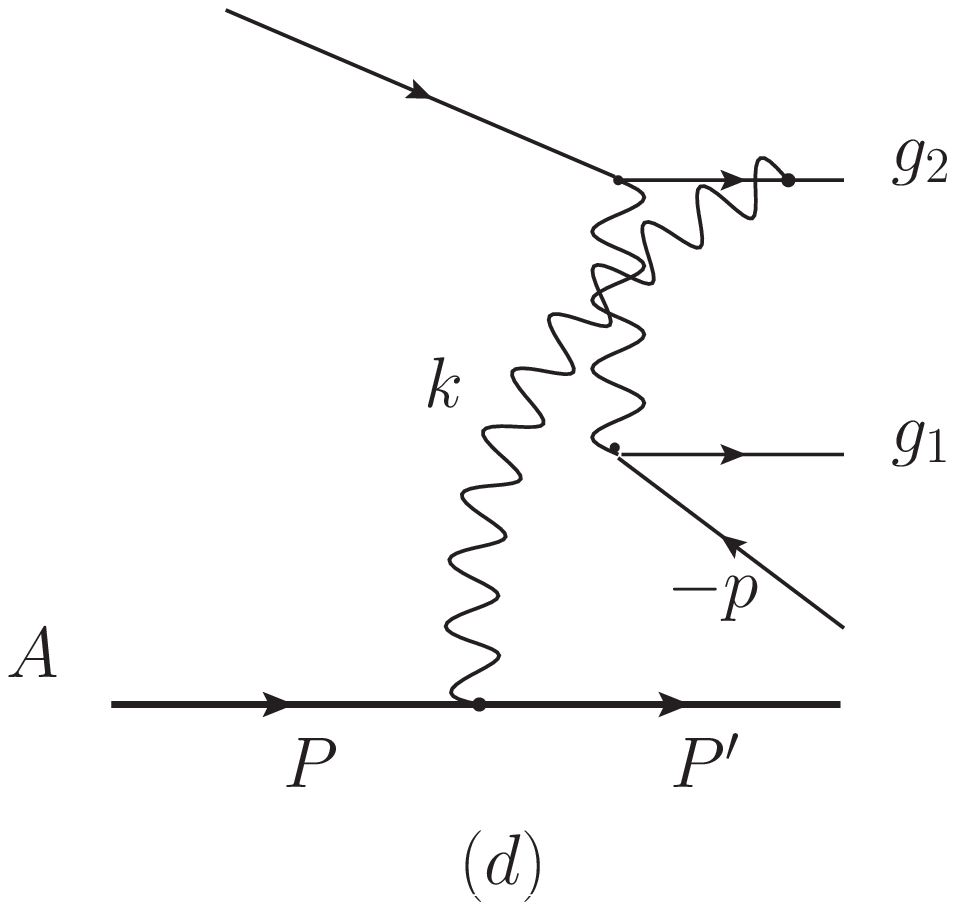}%
\hfill
\end{center}
\caption[*]{Lowest-order graphs for di-jet production in hadron-hadron
collision at small-x limit. In these graphs, there is one soft gluon
exchange with momentum $k$ in addition to the hard gluon exchange.}
\label{fig1}
\end{figure}

We perform our calculations in the covariant gauge.
The final result does not depend on the gauge choice.
We organize the calculations in terms of orders of the
coupling $g$. At each order, a gluon attaches
the scalar quarks in partonic scattering part from the 
nucleus target~\cite{Brodsky:2002ue,Collins:2007nk,{Vogelsang:2007jk}}. 
As shown in Fig.~\ref{fig1}, the lowest-order graphs contain 
one soft gluon exchange with the momentum $k$ and $k^{+}\ll P_{A}^{+}$ 
where the plus component of a momentum $p$ is defined as
$p^{+}=(p^{0}+p^{z})/\sqrt{2}$ and the nucleus is moving in
$+\hat{z}$ direction. We emphasize that the final result is 
frame-independent. We calculate the scattering amplitude in the
infinite momentum frame of the nucleus, i.e., $P_{A}^{+}\rightarrow \infty $.
One can also perform the calculations in the target rest frame and take
the limit of $M_{A}\rightarrow \infty $, which will 
lead to the same result~\cite{Brodsky:2002ue}. 
The small-$x$ approximation ($q^{+}\sim
k^{+}\ll P_{A}^{+}$) will be taken throughout the following
calculations. We will only keep the leading contributions in this
limit. In additional, we also follow the low transverse momentum
approximation in terms of $q_{\perp }/P_{1\perp }$ ($q_{\perp
}/P_{2\perp }$) by applying the power counting
method~\cite{qvy-short}. An important simplification is the
eikonal approximation, which replaces the gluon attachment to the
initial and final state partons with the eikonal propagator and
vertex. After taking the leading order contributions, we find that
the $q_{\perp }$ dependence of these diagrams can be cast into an
effective quark distribution~\cite{qvy-short}, which takes the
following form,
\begin{equation}
\tilde{q}\left( x,q_{\perp }\right) =\frac{x}{32\pi ^{2}}\int \frac{dp^{-}}{%
p^{-}}\frac{d^{2}k_{\perp }}{\left( 2\pi \right) ^{4}}(4P^{+}p^{-})^{2}\left%
\vert A^{(tot)}\left( k,p\right) \right\vert ^{2}\ ,
\end{equation}%
with $p_{\perp }=k_{\perp }-q_{\perp }$. Here the hard
partonic part depending on hard momentum scale $P_{i\perp}$ 
has been separated from the above quark distribution in
the differential cross section~\cite{qvy-short}. This separation
is only possible at the leading power contribution 
of $q_\perp/P_{i\perp}$. 
The contributions from Fig.~2 can be written as,
\begin{equation}
A^{(1)}\left( k,p\right) =gg_{1}\frac{1}{k_{\perp }^{2}+\lambda ^{2}}\left[
\frac{1}{D_{1}}-\frac{1}{D_{2}}\right] \ ,  \label{oneglu}
\end{equation}%
where we have defined $D\left( p_{\perp }\right)
=2xP^{+}p^{-}+p_{\perp
}^{2}+m^{2}$ and $D_{1}=D(q_{\perp })$ and $D_{2}=D(p_{\perp })$. In Eq.~(%
\ref{oneglu}), the first term and second term in the square bracket
correspond to Fig.~\ref{fig1} (a) and Fig.~\ref{fig1} (b), respectively. The
contribution from Fig.~\ref{fig1} (c) and Fig.~\ref{fig1} (d) simply
cancels. That means at the leading order in the coupling constant, the
dependence on $g_{2}$ drops out, which however will change at higher orders.

\begin{figure}[tbp]
\includegraphics[width=4cm]{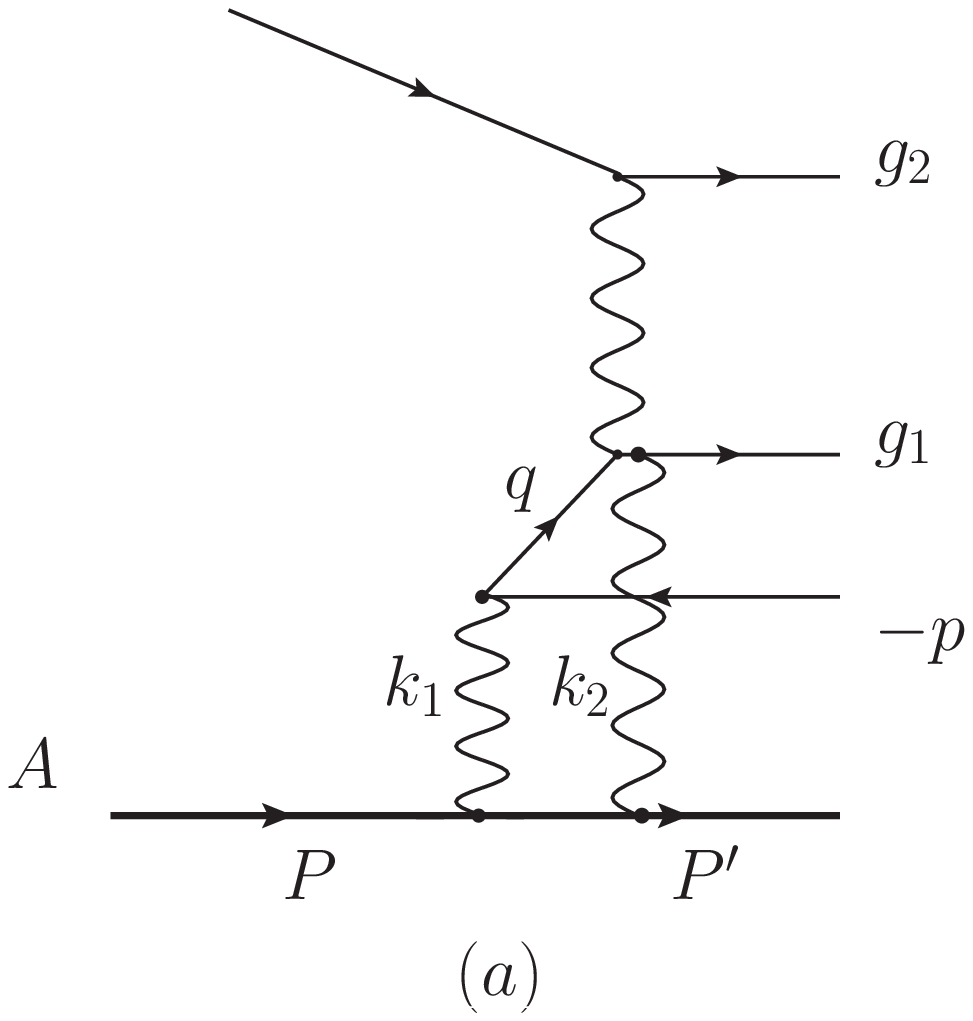}
\includegraphics[width=4cm]{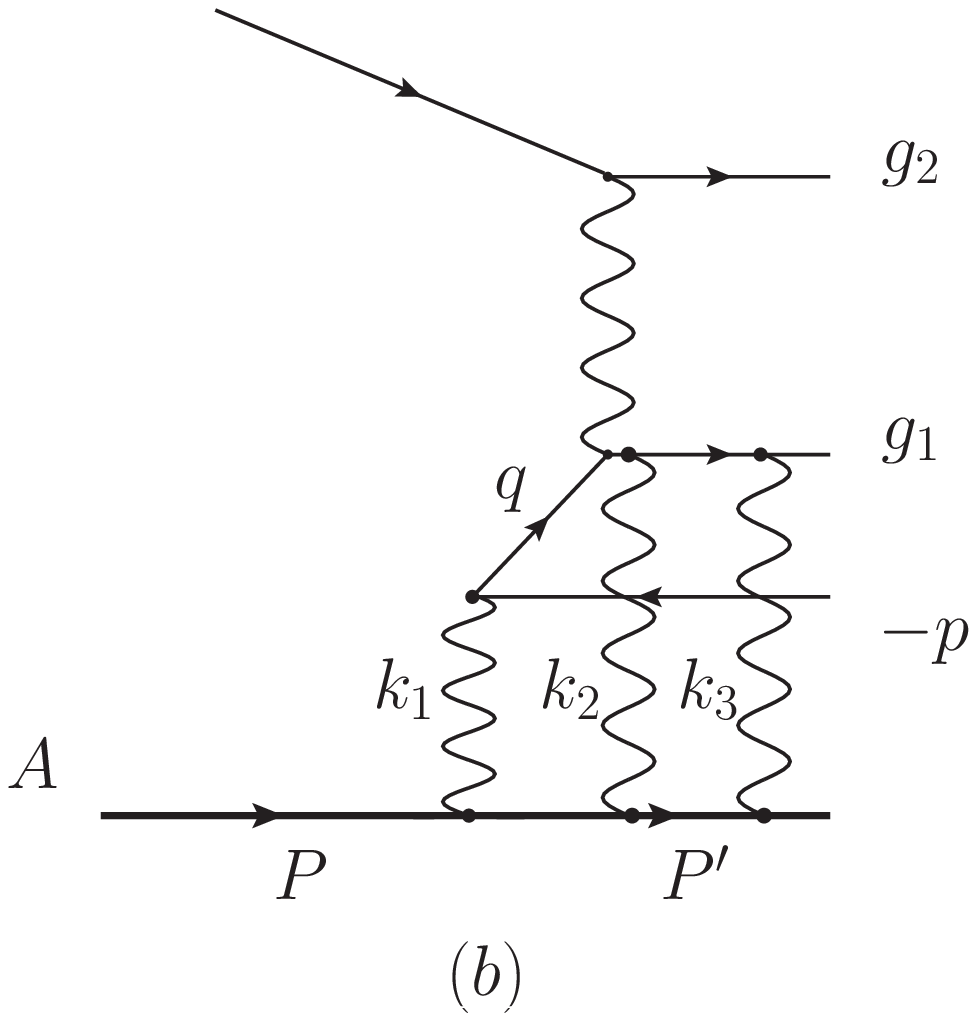}
\caption[*]{Example diagrams for two (a) and three (b) gluons exchanges,
where the gluons can attach all charge particles in the upper part of the
diagrams from the nucleus target.}
\label{fig2}
\end{figure}

At the next-to-leading order, there are 20 graphs in total in covariant
gauge. We plot one of these graphs as an example in Fig.~(\ref{fig2}) (a),
and additional diagrams can be obtained by attaching the gluons to all
incoming and outgoing scalar quarks. The total contributions from these
diagrams are
\begin{eqnarray}
A^{(2)}\left( k,p\right) &=&\frac{i}{2}g^{2}\int d[1]d[2] \left\{g_1^2\left[
\frac{1}{D_1}+\frac{1}{D_2}-\frac{1}{D_{21}}- \frac{1}{D_{22} }\right]
+g_{1}g_{2}\left[ \frac{2}{D_2 }-\frac{2}{D_{21} }\right] \right\}\ ,
\end{eqnarray}
where $\int d[1]d[2]$ stands for $\int \frac{d^{2}k_{1\perp }d^{2}k_{2\perp }%
} {\left( 2\pi \right) ^{4}} \frac{1}{k_{1\perp }^{2}+\lambda^{2}}\frac{1}{%
k_{2\perp }^{2}+\lambda^{2}}(2\pi)^2\delta^{(2)}(k_\perp-k_{1\perp}-k_{2%
\perp})$, $D_{1i}=D(q_\perp-k_{i\perp})$ and $%
D_{2i}=D(p_\perp-k_{i\perp})$. Clearly, this result shows the dependence on $%
g_2$. In order to check this residue dependence in the amplitude squared for
the quark distribution in Eq.~(2), we need to carry out the calculation
of the amplitude up to order $g^3$.

At the $g^{3}$ order, there are 120 diagrams in total with three soft
gluon-exchange (see e.g., Fig.~\ref{fig2} (b)), including all possible
permutation of the attachments of these three gluons on the target nucleus.
Summing up all these graphs, we obtain the three gluon exchange amplitude,
\begin{eqnarray}
A^{(3)}\left( k,p\right) &=&\frac{1}{3!}g^{3}\int
d[1]d[2]d[3]\left\{ g_{1}^{3}\left[ \frac{1}{D_{2}}-\frac{1}{D_{1}}+\frac{3}{%
D_{13}}-\frac{3}{D_{21}}\right] \right.   \notag \\
&&\left. +g_{1}^{2}g_{2}\left[ \frac{3}{D_{2}}+\frac{3}{D_{13}}-\frac{3}{%
D_{21}}-\frac{3}{D_{22}}\right] +g_{1}g_{2}^{2}\left[ \frac{3}{D_{2}}-\frac{3%
}{D_{21}}\right] \right\} \ ,
\end{eqnarray}%
where $\int d[1]d[2]d[3]$ follows similar definition as in Eq.~(4). Again,
we see the dependence on $g_{2}$ coupling in the second and third terms. An
important cross check of these results is that if we set $g_{2}=-g_{1}$,
effectively there is no charge flow in the final state, and the quark
distribution is identical to that in the Drell-Yan process in the same
model. Applying $g_{2}=-g_{1}$, we can easily see that indeed Eqs.~(4,5)
reproduce those calculated in Ref.~\cite{Peigne:2002iw}.

With the amplitude calculated up to $g^3$, we are able to check
the dependence on $g_2$ for the parton distribution in Eq.~(2).
Substituting the results in Eqs.~(3-5) into Eq.~(2), we find that
the $g_2$ dependence still remains up to order $g^4$. If we drop
all $g_2$ terms in these results, we obtain the quark distribution
in deep inelastic scattering process in the same
model~\cite{Brodsky:2002ue,BelJiYua02}. This clearly shows that
the transverse momentum dependent quark distribution $\tilde
q(x,q_\perp)$ is not universal.

This non-universality is better illustrated when we sum up all order
multi-gluon exchange contributions. To do that, we introduce the following
Fourier transform~\cite{Brodsky:2002ue},
\begin{equation}
A\left( R,r\right) =\int \frac{d^{2}k_{\perp }}{\left( 2\pi \right) ^{2}}%
\frac{d^{2}p_{\perp }}{\left( 2\pi \right) ^{2}}e^{-ik_{\perp }\cdot
R_{\perp }-ip_{\perp }\cdot r_{\perp }}A\left( k,p\right) \ .
\end{equation}%
From the Fourier transforms of $A^{(1,2,3)}(k,p)$, we can easily see that
they follow the expansion of an exponential form,
\begin{equation}
A^{(tot)}\left( R,r\right) =\sum_{n=1}^{\infty }A^{\left( n\right) }\left(
R,r\right) =iV\left( r_{\perp }\right) \left\{ 1-e^{igg_{1}\left[ G\left(
R_{\perp }+r_{\perp }\right) -G\left( R_{\perp }\right) \right] }\right\}
e^{-igg_{2}G\left( R_{\perp }\right) }\ ,
\end{equation}%
where $G(R_{\perp })=K_{0}\left( \lambda R_{\perp }\right)/2\pi $ and
$V(r_{\perp })=K_{0}\left( Mr_{\perp }\right) /2\pi $ with
$M^{2}=2xP^{+}p^{-}+m^{2}$ . In the above result, the
$g_{2}$-dependence seems to only appear as a phase which may not lead
to a physics consequence. However, because the transverse
momentum $q_{\perp }$ is conjugate to the coordinate variable difference $%
R_{\perp }$-$r_{\perp }$, this phase will lead to a
non-universality contribution to the quark distribution as defined
in Eq.~(2). Therefore, the all order result reads as,
\begin{eqnarray}
\tilde{q}\left( x,q_{\perp }\right) &\!\!=\!\! &\frac{xP^{+2}}{8\pi ^{4}}%
\int dp^{-}p^{-}\int d^{2}R_{\perp }d^{2}R_{\perp }^{\prime }d^{2}r_{\perp
}e^{iq_{\perp }\cdot \left( R_{\perp }-R_{\perp }^{\prime }\right)
}e^{-igg_{2}\left( G(R_{\perp })-G(R_{\perp }^{\prime })\right) }V\left(
r_{\perp }\right) V\left( r_{\perp }^{\prime }\right)   \notag \\
&&\times \left\{ 1-e^{igg_{1}\left[ G\left( R_{\perp }^{{}}+r_{\perp
}^{{}}\right) -G\left( R_{\perp }\right) \right] }\right\} \left\{
1-e^{-igg_{1}\left[ G\left( R_{\perp }^{\prime }+r_{\perp }^{\prime }\right)
-G\left( R_{\perp }^{\prime }\right) \right] }\right\} \ ,
\end{eqnarray}%
where $r_{\perp }^{\prime }=R_{\perp }+r_{\perp }-R_{\perp }^{\prime }$.
This transverse momentum dependent quark distribution is clearly different
from that calculated in deep inelastic scattering in the 
same model~\cite{Brodsky:2002ue,{BelJiYua02}}. 
In other words, this distribution is not universal. It is
interesting to notice that the $g_{2}$ dependence disappears after the
integration over the transverse momentum. This is consistent with the
universality for the integrated parton 
distributions~\cite{BelJiYua02,{Collins:2007nk},Vogelsang:2007jk}.

\section{Summary and Discussions}

In this paper, we have demonstrated the non-universality for the small-$x$
parton distributions in dijet correlation, by performing
an explicit scalar calculation of the initial/final state interaction effects, and
comparing to those in deep inelastic scattering on the nucleus target. After
summing up to all orders, we find that the net effects are summarized into a
phase which leads to a non-vanishing contribution to the quark distribution
and breaks the universality.

It has been argued that the light-cone gauge may simplify the factorization
property for the hard scattering processes. For example, if we choose the
advanced boundary condition for the gauge potential in light-cone gauge, the
wave function of hadrons contain the final 
state interaction effects~\cite{BelJiYua02,{Brodsky:2010vs}}. However, as we showed in the above
calculations, this does not help to resolve the $g_{2}$-dependence in the
quark distribution in the dijet correlation in hadronic process due to the
presence of both initial and final state interactions. In other words, the
quark distribution of the nucleus in this process has to contain the interaction
with the incoming (outgoing) quark with charge $g_{2}$, which can not be
solely included into the wave function of the nucleus target.

The non-universality effect found in this paper is different from the
factorization breaking effect discussed in Ref.~\cite{Fujii:2005vj}, where
the breaking effect decreases as the heavy quark mass (equivalent to our jet
transverse momentum $P_{1\perp }$) increases. In this paper, we are
discussing the non-universality effects which is in the leading power, and
does not vanish with large transverse momentum of the jet.

Our results indicate that there might be no universality for the transverse
momentum dependent parton distributions at small-$x$. This may impose
a challenge to explain the dijet-correlation data in $dA$
collisions at RHIC with the saturation phenomena observed in DIS
experiments at small-$x$. The non-universality, on the other hand,
provides an opportunity to study the QCD dynamics associated with
the final/initial state interaction effects. These effects
should be taken into account to understand the experimental data.
We plan to address this issue in a saturation model~\cite{Mueller:1999wm,{Mueller:1999yb}} 
in a future publication, together with detailed derivation of this paper.

We thank Les Bland, Stan Brodsky, Paul Hoyer, Larry McLerran, Jianwei Qiu, Raju
Venugopalan and Nu Xu for stimulating discussions. This work was supported in
part by the U.S. Department of Energy under contracts DE-AC02-05CH11231. We
are grateful to RIKEN, Brookhaven National Laboratory and the U.S.
Department of Energy (contract number DE-AC02-98CH10886) for providing the
facilities essential for the completion of this work.

\end{document}